\documentclass[useAMS, usenatbib]{mn2e}
\usepackage{times}
\usepackage{amsmath}
\usepackage{xspace}
\usepackage{psfig}
\usepackage{defs}

\newcommand{\mincir}{\raise
-2.truept\hbox{\rlap{\hbox{$\sim$}}\raise5.truept 
\hbox{$<$}\ }}
\newcommand{\magcir}{\raise
-2.truept\hbox{\rlap{\hbox{$\sim$}}\raise5.truept
\hbox{$>$}\ }}
\newcommand{\minmag}{\raise-2.truept\hbox{\rlap{\hbox{$<$}}\raise
6.truept\hbox
{$>$}\ }}

\renewcommand{\H}{{\mbox{${\rm H{\sc i}~}$}}}

\newcommand{\lya}{Lyman-$\alpha$~}

\newcommand{\eg}{{\it e.g.~}}
\newcommand{\ie}{{\it i.e.~}}

\newcommand{\be}{\begin{equation}}
\newcommand{\ee}{\end{equation}}
\newcommand{\ba}{\begin{eqnarray}}
\newcommand{\ea}{\end{eqnarray}}
\newcommand{\brr}{\begin{array}}
 
\newcommand{\err}{\end{array}}
\newcommand{\bc}{\begin{center}}
\newcommand{\ec}{\end{center}}

\newcommand{\fu}{\,{\rm erg} \, {\rm cm}^{-2} \, {\rm s}^{-1}}
\newcommand{\vel}{\,{\rm km\,s^{-1}}}

\newcommand{\yr}{{\, \rm yr} }
\newcommand{\msun}{\, M_{\odot }}
\newcommand{\dd}{\, {\rm d}}

\newcommand{\ltsima}{\mbox{$\; \buildrel < \over \sim \;$}}

\def\eq{equation}
\def\fig{Fig.}

\def\ie{{\it i.e.}}
\def\eg{{\it e.g.}}
\def\ltsima{$\; \buildrel < \over \sim \;$}
\def\simlt{\lower.5ex\hbox{\ltsima}}
\def\gtsima{$\; \buildrel > \over \sim \;$}
\def\simgt{\lower.5ex\hbox{\gtsima}}
\def\eq{equation}
\def\H2{H$_2$\xspace} 
\def\m{$^{-1}$\xspace}

\def\h2{H$_2$\xspace}
\def\ion#1#2{\text{#1\,\sc #2}}
\def\HI{{\ion{H}{i} }}

\def\popII{Population~II\xspace}
\def\pop3{Population~III\xspace}
\def\Mpc{$h^{-1}$~Mpc\xspace}

\def\Lya{Ly$\alpha$\xspace}
\def\xhI{$\langle x_\HI \rangle_M$\xspace}

\def\sig8{$\sigma_8$\xspace}

\def\bi{\begin{itemize}}
\def\ei{\end{itemize}}

\DeclareMathAlphabet{\mathsc}{OT1}{cmr}{m}{sc}
\def\testbx{bx}%
\DeclareRobustCommand{\ion}[2]{%
\relax\ifmmode
\ifx\testbx\f@series
{\mathbf{#1\,\mathsc{#2}}}\else
{\mathrm{#1\,\mathsc{#2}}}\fi
\else\textup{#1\,{\mdseries\textsc{#2}}}%
\fi}

\title[Have we detected one of the sources responsible for an early 
reionisation of hydrogen?]{Have we detected one of the sources responsible
for an early reionisation of the Universe?}
\author[Massimo Ricotti, Martin G. Haehnelt, Max Pettini and Martin J. Rees]
{Massimo Ricotti, Martin G. Haehnelt, Max Pettini and Martin J. Rees \\ 
$^1$ Institute of Astronomy, Madingley Road, Cambridge CB3 0HA\\
\\}

\begin{document}

\maketitle
\begin{abstract}
  In a recent paper Pell\'o et al. have reported a candidate $z=10$
  galaxy, A1835\#1916, which was found in a near-infrared survey of
  the central regions of the gravitational lensing cluster A 1835. If
  this detection is confirmed and the detection rate turns out to be
  typical, then the volume averaged ultraviolet emissivity must be
  rising rapidly with increasing redshift. For a magnification due to
  gravitational lensing by a factor ${\cal M}\ga 25$ estimated by
  Pell\'o et al., the inferred star formation rate at $z=10$ would be
  about one order of magnitude higher than estimates of the star
  formation rate density at $z=6$.  Objects at $z=10$ would
  contribute substantially to the total source counts at $1.6 \mu {\rm
    m}$ and the estimated space density of sources may exceed the
  space density of dark matter haloes in a $\Lambda$CDM model.  We
  therefore argue that if A1835\#1916 is indeed at $z=10$ then either
  the magnification factor may have been overestimated or the galaxy
  has a top-heavy initial mass function.  Sources with the UV flux and
  space density of A1835\#1916 produce $\sim 33 f_{\rm esc} ({\cal
    M}/25)$ hydrogen ionising photons per hydrogen atom per Hubble
  time, where $f_{\rm esc}$ is the escape fraction of ionising
  photons.  This rate should be sufficient to reionise most of the
  diffuse hydrogen in the Universe at redshift ten.  We further use a
  correlation between the equivalent width and the redshift of the
  \Lya emission line with respect to the systemic redshift observed in
  Lyman break galaxies to obtain constraints on the ionisation state
  of the surrounding intergalactic medium (IGM) from the Gunn-Peterson
  absorption. These constraints also argue in favour of the
  surrounding IGM being fully ionised.  Pell\'o et al.  may thus have
  detected a population of sources which is responsibly for the large
  electron scattering optical depth indicated by the cross-power
  spectrum of the temperature and polarisation fluctuations of the
  cosmic microwave background as measured by WMAP.
\end{abstract}


\section{Introduction}
The unravelling of the reionisation history of the Universe has been
the focus of much recent research mainly due to the surprising
detection of a large Thompson electron optical depth of $\tau =0.17\pm
0.04$ by WMAP \citep{Kogut:03}.  If correct, this optical depth
requires a substantial ionised fraction of hydrogen at redshift
$z=10-20$. This result came somewhat as a surprise as the optical
depth for \Lya scattering increases rapidly in the highest redshift
QSOs \citep{Fan:03, Songaila:04} indicating a drop in the emissivity
of hydrogen ionising photons \citep{Miralda:03b}. This has led to the
suggestion that the emissivity of ionising photons peaked at high
redshift due to a population of early stars or mini-AGN and that the
reionisation may have been complicated, with an extended epoch of
partial reionisation \citep{MadauR:03, RicottiO:03} and/or the
possibility that hydrogen was reionised twice \citep[\eg,][]{Cen:03a,
  WyitheL:03, CiardiFW:03, RicottiOI:03,Sokasian:03}. The possible
detection of a $z=10$ \Lya emitting star-forming galaxy by
\cite{Pello:04} thus offers exciting prospects to further constrain
the reionisation history.  This result, if confirmed by further
observations, pushes back the epoch when ``galaxies'' were already
shining in Ly$\alpha$ emission by $\sim 0.5$\,Gyr relative to previous
detections at $z \simeq 5 - 6$ \citep{Rhoads:03, Kodaira:03,
  Lehnert:03, Hu:04, Stanway:04}. In this letter, we briefly discuss
the implications of detecting such a source for the inferred space
density of star-forming galaxies, for the emissivity of hydrogen
ionising photons and the ionisation state of the IGM at $z=10$.  We
assume throughout the cosmology to be the concordance $\Lambda$ cold
dark matter model with ($\Omega_\Lambda, \Omega_{\rm m}, \Omega_{\rm
  b}, {\rm h}) = (0.7, 0.3, 0.04, 0.7$) and primordial scale-invariant
power spectrum ($n_s=1$) with rms amplitude of the mass fluctuations
on scale of 8 \Mpc, $\sigma_8=0.91$.

\section{The $z=10$ candidate galaxy A1835\#1916}

\subsection{Summary of observations}\label{sec:obs}
Pell\'o et al. (2004, P04) obtained deep ISAAC imaging in JHK of the
central $2\times 2 {\rm \, arcmin}^2$ of the gravitational lensing
cluster A1835. Together with deep optical imaging in VRI they were
able to identify 6 high-redshift ($z>7$) candidates using the dropout
technique \citep{Guhathakurta:90, Steidel:92}.  One of the candidates
(\#1916) has a redshift estimate from broad-band photometry of $z_{\rm
  phot} \approx 9-11$ and falls close to the critical line of the
cluster for this redshift range.  For this candidate P04 obtained a
deep J-band spectrum with ISAAC and detected an emission line at
1.33745 $\mu {\rm m}$ with a flux of $4 \times 10^{-18} \fu$ and a
rest frame width of $\approx 50 \vel$.  If interpreted as \Lya
emission the redshift of the emission line is $z= 10.00175$. From the
location of the source relative to the critical lines of their lensing
model, P04 estimate the amplification due to gravitational lensing to
be in the range $25 <$ $ \cal{M}$ $< 100$.  The inferred star
formation rates (uncorrected for lensing) from the line and continuum
fluxes are $ 4 h_{70}^{-2} \msun \yr^{-1}$ and $60 h_{70}^{-2} \,
\msun \yr^{-1}$ respectively, using the observed fluxes reported by
P04 and the conversions by \citep{Kennicutt:98}. Note that P04 used a
different conversion from \Lya flux to star formation rate, obtaining
a lower value of the star formation rate.

%

\subsection{The implied space density of star-forming galaxies and
star formation rate density at z=10}\label{sec:stat}

The comoving survey volume per unit redshift is given by
\begin{equation} 
\frac{d V_{\rm survey}(z)}{\dd z} = \frac{c^3}{\cal M}  
\left ( \int_0^{z}{\frac{1}{H(z')}}  d z' \right ) ^2
\frac{1}{H(z)} \dd \Omega,
\end{equation}
where $\cal{M}$ is the magnification by gravitational lensing and 
$\dd \Omega$ is the solid angle of the area surveyed in the lens plane.
  
The solid angle in the lens plane with magnification larger than a
given value is somewhat uncertain and depends on the lens model.  The
length of the critical curve in Fig.~1 of P04 is about 180 arcsec.  We
further assume that for sources within 2.5 arcsec of the critical
curve ${\cal M} >25 $, then $\dd \Omega(> {\cal M}) \approx 0.25
({\cal M}/25)^{-1} {\rm arcmin}^2$ (R. Pello private comunication).
This is about the same value as \cite{SantosE:03} give in their Fig.~7
as average for 9 lensing clusters and assuming that the slits cover
about 1/5 of the total magnified area.

Assuming that the survey detects all galaxies in the redshift range
$8.5<z<10.5$, and that the detection of one galaxy per effective survey
volume is representative, the number density of bursting sources with
star formation rate ${\rm SFR} \ga 2.4 ({\cal M}/25)^{-1} h_{70}^{-2} \msun
\yr^{-1}$ is given by,
\begin{equation}
\begin{split} 
n_{\rm burst} \approx
             \left( \frac{d V_{\rm survey}}{\dd z} 
                 \Delta z   \right)^{-1} \\ \simeq 0.033
\left ( \frac{\dd \Omega_{\rm eff}}{0.25~{\rm arcmin}^2   } \right )^{-1}
\left ( \frac{\cal M}{25} \right )^{2}
h_{70}^{3} \, {\rm Mpc}^{-3} ,
\end{split}
\label{eq:nburst}
\end{equation} 
where $\dd \Omega_{\rm eff}({\cal M}) $ is the effective solid angle
in which such a source with magnification factor $>{\cal M}$ is {\it
  typically} found.  This number is obviously very uncertain as P04
have just found one source and the magnification is also uncertain.
However, as we will show later on, despite this large uncertainty the
high value of the estimated galaxy space density has interesting
implications.  Note that the total space density of sources, $n_{\rm
  tot}$, including those not currently undergoing a starburst is a
factor $10 (t_{\rm burst} / 30\, {\rm Myr})^{-1}$ larger than that in
\eq~(\ref{eq:nburst}).  Note also that this would correspond to $3.6
\times 10^5 (\dd \Omega_{\rm eff}/0.25 )({\cal M}/25)^{2}$ objects per
${\rm deg}^2$ with H-band AB magnitude of $28.5 + 2.5 \log ({\cal
  M}/25)$ which for magnification factors in the range suggested by
P04 approaches the observed number counts at $1.6 \mu {\rm m}$
\citep{Yan:98, Thompson:99}.



\begin{figure}
\centerline{\psfig{figure=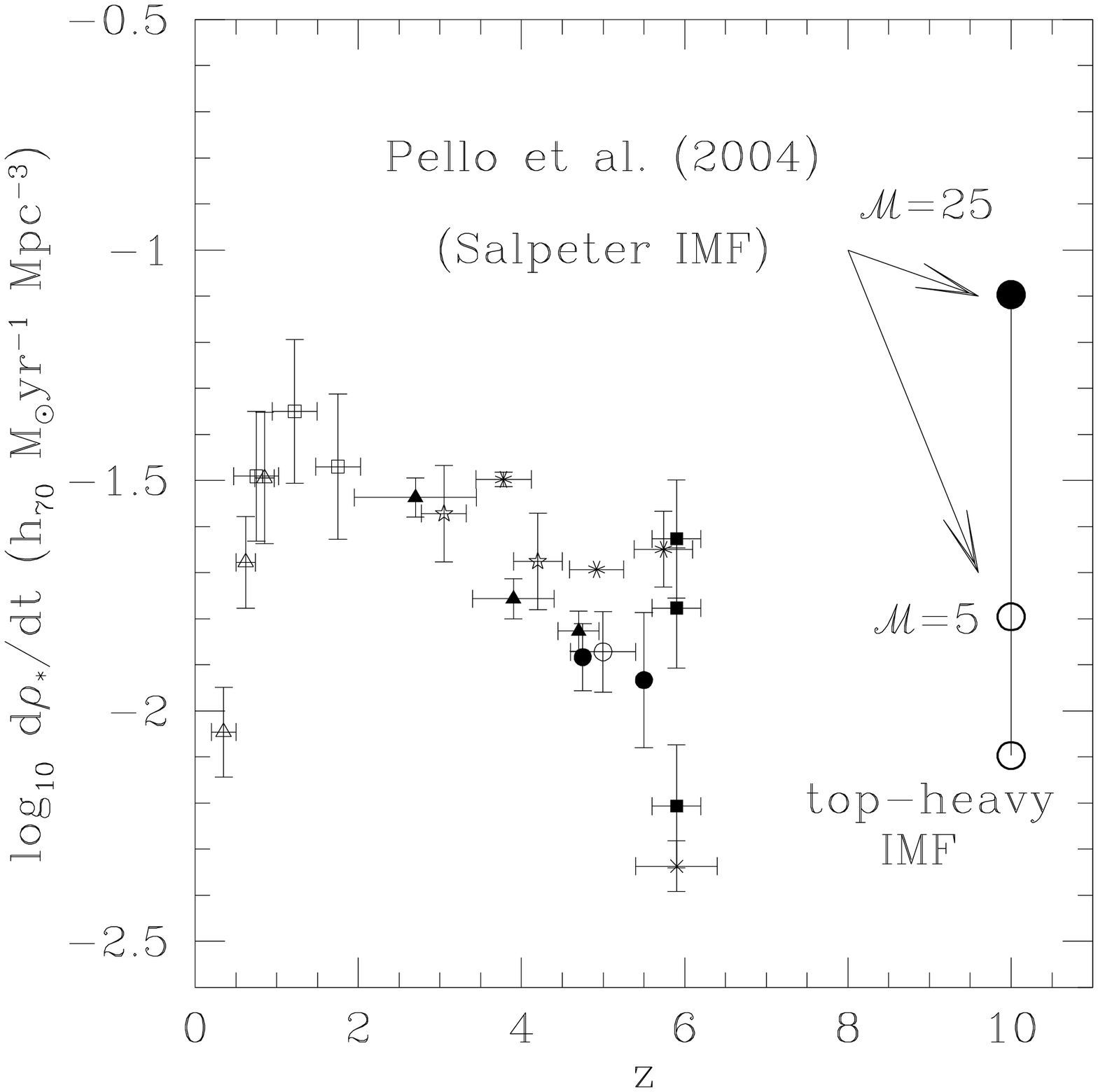,width=8.5cm}}
\caption{\label{fig:madau} The star formation rate density as
  a function of redshift adopted from \protect\cite{Bunker:04}.  The
  data is from \protect\cite{Lilly:96} (open triangles),
  \protect\cite{Connolly:97} (open squares), \protect\cite{Steidel:99}
  (stars), \protect\cite{Iwata:03} (open circles),
  \protect\cite{BouwensBI:03} (solid triangles),
  \protect\cite{Bouwens:03} (solid squares), \protect\cite{Fontana:03}
  (solid circles), \protect\cite{Giavalisco:04} (asterisks),
  \protect\cite{Bunker:04} (crosses). The value inferred from Pell\'o
  et al. (assuming a Salpeter IMF) is shown as a solid circle for our
  fiducial values of the magnification (${\cal M} = 25$) and the
  estimated volume of the survey.  The point in the middle of the
  error bar (open circle) is obtained assuming five times smaller
  magnification of the source and the point at the end of the error
  bar assuming a top-heavy IMF.}
\end{figure}

Taken at face value, the space density is about a factor $(0.6 -
2)({\cal M}/25)$ times that of \Lya emitters in surveys at redshift 4
to 6 \citep[\eg,][]{SantosE:03}.  It corresponds to a star formation
rate density of $\rho_{*} \approx 0.08 \, (\dd \Omega_{\rm eff}/0.25
{\rm arcmin}^2 )^{-1} ({\cal M}/25) \, h_{70} \msun \yr ^{-1}\, {\rm
  Mpc}^{-3} $.  The total star formation rate is expected to be larger
due to the contribution of fainter objects with smaller star formation
rates.  In \fig~\ref{fig:madau} we compare this star formation rate to
the compilation of star formation rates at lower redshifts by
\cite{Bunker:04}.  The star formation rate density for our fiducial
values is about a factor 4-20 larger than that observed at redshift
six \citep{Bouwens:04,Bunker:04}. We also show the inferred star
formation rate density assuming ${\cal M}=5$ and a top-heavy initial
mass function (IMF) (middle and lower point on the error bar
respectively).  If A1835\#1916 is indeed at $z=10$, ${\cal M}=25$ and
the average volume that contains such a source is not underestimated
the ultraviolet emission rate density would have to increase rather
rapidly.  Such a rapid rise of the emissivity towards larger redshift
may explain why the detected source has not been found closer to the
lower end of the redshift range where it could have been detected
($z\sim 7 -8$), as it is most likely in a flux-limited sample. A
continuation of the decrease of the comoving star formation rate
density between $z=4$ and $z=6$ suggested by Bunker et al. (2004)
would only be possible if the magnification factor and/or the space
density of the sources have been overestimated, or if the IMF at $z
\sim 10$ becomes top-heavy.

\subsection{The space density of galaxies and DM haloes at z=10}\label{sec:mag}

As discussed in the last section the inferred space density of sources
and star formation rate density are quite large for a Salpeter IMF if the
magnification $25 < {\cal M }< 100$ derived by P04 from the location
of the source relative to the critical lines of their lens model is
correct.

\begin{figure}
\centerline{\psfig{figure=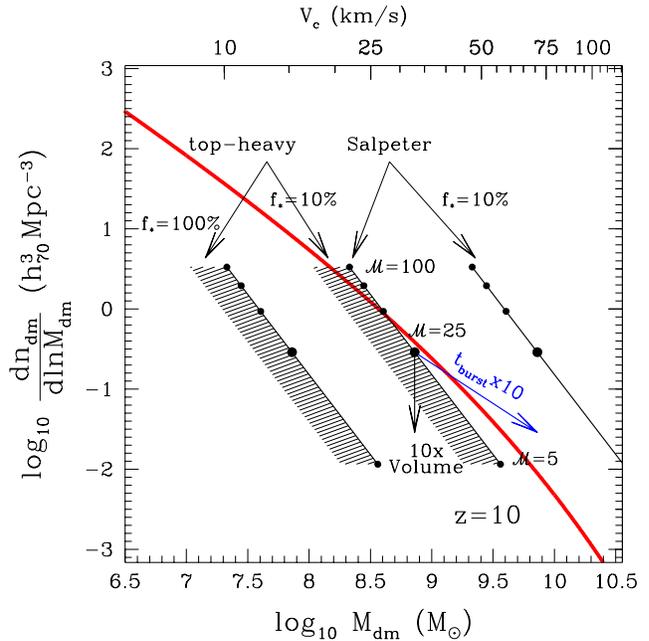,width=8.5cm}}
\caption{\label{fig:nps} The solid curves shows the mass function of
  dark matter haloes at $z=10$ for a $\Lambda$CDM concordance model.
  The hatched regions show lower limits for the total space density of
  haloes hosting sources like A1835\#1916 for a range of magnification
  and two different IMF. We have assumed that 100\% or 10\% (as
  indicated by the labels) of all available baryons turn into stars
  within $t_{\rm burst}=30$ Myr. The arrows show how the limits change
  with increased duration of the burst and increased typical volume in
  which a source like A1835\#1916 can be found.}
\end{figure}

To investigate this further we show in \fig~\ref{fig:nps} the expected
space density of dark matter haloes in the concordance $\Lambda$CDM
model at redshift $z=10$ (thick solid curve).

In order to compare that to the space density of the observed
source(s) we have to assume a duration of the burst, a star formation
efficiency and an IMF.  The total number of haloes required to host
the bursts is $n_{\rm tot} \simeq 0.32 (t_{\rm burst} / 30\, {\rm
  Myr})^{-1} \, ( \Omega_{\rm eff}/0.25\, {\rm arcmin}^2 ) \,({\cal
  M}/25)^2$ h$_{70}^3$ Mpc$^{-3}$.

We will start by assuming that all baryons in a DM halo turn into
stars (\ie, we assume a star formation efficiency $f_*=100\%$) on a
time scale $t_{\rm burst}$ so that $\dot M_{*} = 0.14 M_{\rm dm} /
t_{\rm burst}$.  This gives a (lower limit of the) mass of the dark
matter halo hosting A1835\#1916 of $M_{\rm dm} \simgt 4.2 \times
10^8(t_{\rm burst}/30~ {\rm Myr})(25/{\cal M})$ M$_\odot$, if we
assume a Salpeter IMF. For a top-heavy IMF the inferred mass would be
a factor up to ten smaller. Note that the  galaxy mass function is
expected to have shallower slope at small masses when feedback effects
are included.

The hatched regions show the resulting lower limits of space density
for a range of magnifications from $5-100$ as indicated on the figure.
The arrows show how these limits would change if the burst duration is
increased by a factor ten or the assumed typical volume hosting such a
source has been underestimated by a factor of ten.  Independent of our
detailed assumptions, sources with such a high space density must be
hosted in rather shallow potential wells with virial velocities
$v_{\rm vir} \la 50 \vel$ which fits in well with the narrow width of
the \Lya emission line.  The star formation efficiency in shallow
potential wells ($M_{\rm dm}<10^7-10^8$ M$_\odot$) is,  however,
expected to be of order 10\% \citep[\eg,][]{RicottiGSb:02}. 
The corresponding increase of the estimates for the halo masses is
also  shown in \fig~\ref{fig:nps}.

For a Salpeter IMF and 10\% star formation efficiency our estimated
space density substantially exceeds the simple $\Lambda$CDM
prediction.  However, for a top-heavy IMF both the assumed star
formation efficiencies agree with the model. If our assumed estimates
of the space density and/or the magnification factor are somewhat too
large, then even assuming a shallower galaxy mass function (expected
if feedback effects are important) would still agree with the
$\Lambda$CDM prediction and with a star formation efficiency as low as
10\%.  As discussed above the space density is obviously very
uncertain and a more moderate amplification of say ${\cal M}\sim 5$
may actually not be implausible given that there will be uncertainties
in the model of the gravitational lens and the assumed cosmological
model.  The shallower slope of the galaxy mass with respect to the
halo mass function that is expected when feedback effects are included
also suggests that a large magnification is unlikely.


\section{The ionisation state of hydrogen at z=10}\label{sec:xh}

The observed candidate galaxy at $z=10$ offers two routes to
constraining the ionisation state of hydrogen in the IGM.  With the
star formation density from section~\ref{sec:stat} we can estimate the
total ionising emissivity, while the observed \Lya emission line and
its equivalent width can constrain the Gunn-Peterson optical depth. We
now discuss each of these constraints in turn.

\subsection{The ionising emissivity} 

For a Salpeter IMF $\sim 4000$ ionising photons are produced per
hydrogen atom in the matter turning into stars
\citep[\eg\,][]{Haiman:02}. The emissivity of hydrogen ionising
photons per hydrogen atom per Hubble time, $t_H$, is then $t_H \dd
(n_{\gamma}/n_{H})/\dd t \sim 33 h_{70} f_{\rm esc} ( \dd \Omega_{\rm
  eff}/0.25 {\rm arcmin}^2 )^{-1} ({\cal M}/25) $, where $f_{\rm esc}$
is the escape fraction of ionising photons.  Recall that this is only
the contribution from objects bright enough to be detected. Note also
that for population III stars the emissivity could be larger by a
factor up to two \citep[\eg,][]{Tumlinson:00, Bromm:01}.

It is somewhat uncertain how many photons are needed to actually
reionise the Universe and estimates vary from a total of a few to a
few tens of photons per hydrogen atom \citep{MadauHR:99, Miralda:00,
  Haiman:01}.  For a magnification of ${\cal M}=25$ the emissivity of
ionising photons should be sufficient to fully reionise hydrogen.

\subsection{Suppression of the \Lya emission due to Gunn-Peterson absorption} 

\begin{figure}
\centerline{\psfig{figure=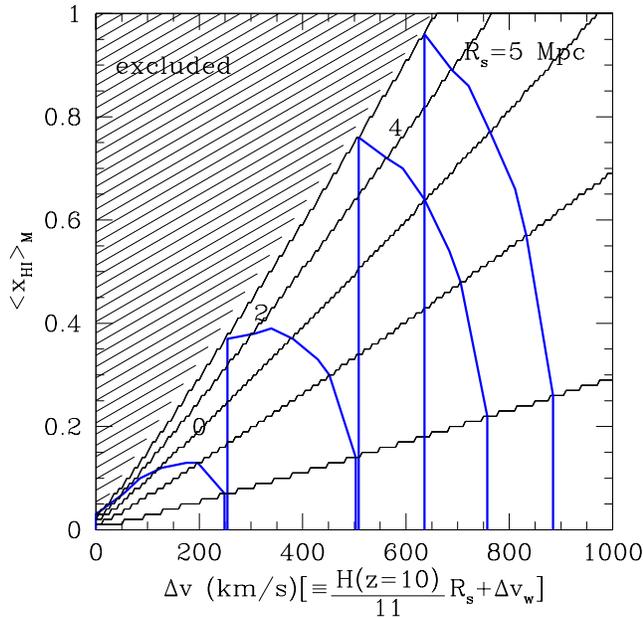,width=8.5cm}}
\caption{\label{fig:igm} The thin solid curves show upper limits on the 
  (mass-weighted) neutral fraction of hydrogen as a function of the
  redshift $\Delta v = H(z) R_{\rm S}/11 + \Delta v_{\rm w}$ relative
  to the absorbing gas just outside the Str\"omgren sphere, where
  $H(z)$ is the Hubble constant, $R_{\rm S}$ is the (comoving) radius
  of the Str\"omgren sphere in which the source is embedded and
  $\Delta v_{\rm w}$ is the peculiar velocity of the emitting gas with
  respect of the Hubble flow . The limits were calculated assuming
  that the absorption of the \Lya emission inferred from the
  equivalent width of the \Lya line is due to combined absorption of
  the surrounding IGM and some intrinsic absorption by the ISM,
  galactic winds and gas inside the Str\"omgren sphere surrounding the
  source.  The thin solid curves are for transmission factors due to
  intrinsic absorption of $T_{\rm w}=100, 67, 42, 22$, and 10\% ,
  respectively (top to bottom).  The thick solid curves show the upper
  limits of the neutral hydrogen fraction assuming that the source is
  at the centre of a Str\"omgren sphere of comoving radius $R_{\rm
    S}=0, 2, 4$, and $5$ Mpc (from left to right) assuming the
  correlation of $T_{\rm w}$ with $\Delta v_{\rm w}$ shown in
  \fig~\ref{fig:dv}.}
\end{figure}

The line profile appears not to show the characteristic asymmetry due
absorption by the surrounding IGM/ISM seen in typical high-redshift
\lya emitters. However, considering the very low S/N this is probably
not a reason for concern.  The P04 estimate for the star formation
rate based on the \Lya emission is a factor 15 smaller than that based
on the UV continuum emission, suggesting that \Lya is strongly
absorbed either by absorption intrinsic to the source or due to the
\Lya opacity of the IGM in front of the source \citep{Miralda:98}, or
both.  We can therefore write the observed \Lya emission as $I_{\rm
  obs} = T_{\rm w} T_{\rm IGM} I_{\rm em}$ where $T_{\rm w}$ and
$T_{\rm IGM}$ are the transmission factors for absorption by the IGM
and intrinsic absorption, respectively, and $T_{\rm w} T_{\rm IGM}
\approx 0.067$.  The transmission of the IGM is related to the optical
depth of the red wing of the \Lya absorption trough produced by the
IGM in front of the source as $T_{\rm IGM} = 1 -\exp{(-\tau_{\rm
    IGM})}$. The IGM optical depth $\tau_{\rm IGM}$ will depend on the
(comoving) radius $R_{\rm S}$ of the Str\"omgren sphere in which the
source is embedded and the peculiar velocity of the emitting gas
$\Delta v_{\rm w}$ with respect of the Hubble flow \citep{Haiman:02,
  Santos:03}.  The emitting gas then has a redshift $\Delta v = H(z)
R_{\rm S}/11 + \Delta v_{\rm w}$ relative to the absorbing gas just
outside the Str\"omgren sphere, where $H(z)$ is the Hubble constant.
If we further assume that the absorbing IGM outside the Str\"omgren
sphere has no peculiar velocity relative to the Hubble flow, the
opacity is given by
\[
\tau_{\rm IGM}(\Delta v) = c\int_{10-\Delta z}^{10} n_{\rm
  HI}(z)\sigma_{Ly\alpha} (\Delta v){dt \over dz}dz,
\]
where $\sigma_{Ly\alpha}$ is the cross section for \Lya absorption and
$n_{\rm HI}$ is the number density of neutral hydrogen.  The integral
converges for $\Delta z \ga 1$.  We do not know the relative
contribution of intrinsic and IGM absorption to the total
transmission.  In \fig~\ref{fig:igm} we therefore show the upper limit
on the mean (mass weighted) neutral fraction, \xhI, of the IGM as a
function of $\Delta v$ for a range of values of the intrinsic
transmission $T_{\rm w}$.

As expected for small values of $\Delta v$, \ie, a small $R_{\rm S}$
and a small $\Delta v_{\rm w}$, the surrounding IGM would have to be
fully ionised.  Otherwise the \Lya emission would be completely
absorbed by the red wing of the Gunn-Peterson trough.  If there was no
intrinsic absorption (\ie, $T_{\rm w} =1, T_{\rm IGM} = 0.067$) a
$\Delta v [\equiv H(z=10) R_{\rm S}/11 + \Delta v_{\rm w}] = 650 \vel$
would be required to be consistent with a fully neutral surrounding
IGM.

The constraints on the neutral state become considerably stronger if
we allow for a significant fraction of the absorption to be intrinsic.
If the intrinsic absorption is 90\%, (\ie, $T_{\rm w} =0.1, T_{\rm
  IGM} = 0.67$) the neutral fraction of the IGM would have to be
smaller than 20\% for values of $\Delta v$ as large as $1000 \vel$.
It would thus help greatly if we could put some constraint on range of
plausible values of $T_{\rm w}$.

\begin{figure}
\centerline{\psfig{figure=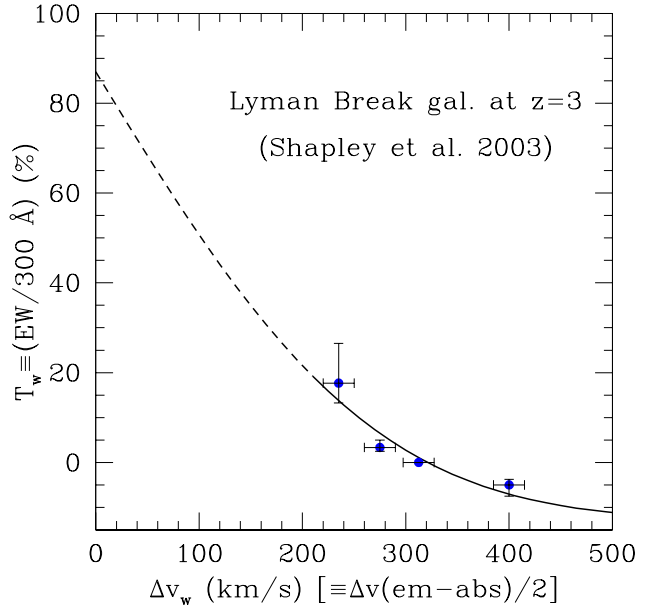,width=8.5cm}}
\caption{\label{fig:dv} Observed correlation between the \Lya
  equivalent width, and thus the transmission factor, with the
  velocity shift of the \Lya emission line with respect to the rest
  frame velocity of the galaxy. The solid curve shows a fit, motivated
  in the text. The dashed line shows the extrapolation to small
  $\Delta v_{\rm w}$. Note that the upper limits on \xhI in
  \fig~\ref{fig:igm} are not sensitive to the particular choice of the
  extrapolation and the assumed typical wind velocity $\sigma_{\rm
    w}$.}
\end{figure}

An approximate estimate can be obtained from studies of Lyman break
galaxies (LBGs) at redshift $z=3-4$ \citep[\eg,][]{Shapley:03}. LBGs
show \Lya either in absorption or emission. For LBGs with \Lya
emission, there is a wide range of equivalent widths and the centre of
the \Lya emission line is generally redshifted by $\approx 200-300
\vel$ relative to the stellar absorption lines and nebular emission
lines which presumably are at the systemic redshift of the galaxy
\citep{Shapley:03}.  This systematic offset is generally taken as
evidence for galactic winds and the \Lya emission is believed to come
from outflowing matter on the far side of the galaxy. Interestingly,
\cite{Shapley:03} find a correlation between the equivalent width (EW)
of the \Lya line and its velocity shift, which we have reproduced in
\fig~\ref{fig:dv}.  The unabsorbed equivalent width (EW$_0$) is
determined by the age, IMF and metallicity of the stellar population
producing it.  Typical values are in the range EW$_0=240-350$~\AA~for
\popII stars and EW$_0=400-850$ \AA~for \pop3 stars
\citep{Schaerer:03}.  If the line is absorbed by local gas and by the
the wind, the EW will be reduced by a factor $T_{\rm w}$. The measured
\Lya equivalent widths at $z \sim 3$ should thus be a good proxy for
the intrinsic transmission of LBGs.  If the profile of the unabsorbed
\Lya line is a Gaussian with $\sigma_{\rm w}=300$ km s\m and part of
the blue wing of the line is absorbed by a galactic wind, it should be
possible to approximate the correlation by $T_{\rm w}={\rm
  Erfc}(\Delta v_{\rm w}/\sigma_{\rm w})$, where Erfc(x) is the
complementary error function and $T_{\rm w}={\rm EW/EW}_0$. We indeed
obtain a reasonable fit shown as the solid and dashed curve in
\fig~\ref{fig:dv} with EW$_0=(300\pm 100)$~\AA~\citep{Schaerer:03},
and a small offset of 13\% of $T_{\rm w}$.

If we assume that the inferred correlation between intrinsic
transmission and redshift of the emitting gas relative to the systemic
redshift found at $z \sim 3$ for LBGs also holds for A1835\#1916, we
can specify the intrinsic transmission for a fixed size of the
Str\"omgren sphere.  The thick solid curves in \fig~\ref{fig:igm} show
these significantly tighter constraints for a range of radii of the
Str\"omgren sphere. We have thereby assumed a wind velocity
$\sigma_w=300$ km s\m, but the upper limits do not change if we assume
$\sigma_w=100$ km s\m which may be more appropriate for sources hosted
in shallow potential wells as is likely for A1835\#1916. The curves
also depend only very weakly on the assumed extrapolation of the
correlation of $T_{\rm w}$ with $\Delta v_{\rm w}$ towards small
velocities.  For $R_{\rm S}\la 5$ Mpc (comoving) the surrounding IGM
must be at least partially ionised to be consistent with the observed
\Lya emission if the $T_{\rm w} - \Delta v_{\rm w}$ correlation of
LBGs holds for A1835\#1916. 

\section{Discussion and Conclusions}

The implications of the possible detection of a redshift 10 galaxy by
\cite{Pello:04} depend crucially on the assumed magnification factor
and on the assumption that the detection of one source of this kind in
the effective survey volume is representative. Our estimate of the
space density is consistent with the predicted number densities of DM
haloes in a LCDM model. For a Salpeter IMF, it requires, however, a
rather extreme (close to $100\%$) star formation efficiency. For a
``top-heavy'' IMF a star formation efficiency $\ge 10\%$ would be
sufficient. If stellar feedback is important in these objects then the
slope of their mass function is expected to be shallower than the mass
function of DM haloes in a $\Lambda$CDM model. A somewhat smaller
magnification and/or space density would then be required to be
consistent with the model prediction. It seems thus worthwhile to
investigate if a smaller magnification is consistent with the
uncertainties in the model of the gravitational lens and the assumed
cosmological model.  If the magnification was indeed smaller, then the
location of the yet missing counter-image would also be much less well
constrained.  More objects of this kind are clearly needed for a more
solid assessment of their space density and the implied emissivity.  A
large star formation rate density and a moderate amplification factor
of this source would obviously be good news for ongoing surveys for
objects at $z>7$, both behind lensing clusters and in the field.  For
the fiducial magnification ${\cal M} \sim 25$ and assuming Salpeter
IMF, the evolution of the star formation rate density would show a
very rapid decrease between $z=10$ and $z=6$ followed by a increase
between $z=6$ and $z=4$ which would suggest that A1835\#1916 is part
of a separate population of objects. This raises the question what
could have led to the rapid decline of such a population.  A number of
authors have made the suggestion that negative feedback due to star
formation in shallow potential wells could lead to a rapid decline of
a first generation of star-forming objects
\citep[\eg,][]{Efstathiou:92b, Barkana:00, Ricotti:02}. If
instead the IMF becomes top-heavy at $z \simgt 10$, then the star
formation rate density is consistent with the decreasing values
observed between $z=4$ and $z=6$. For the fiducial magnification
${\cal M} \sim 25$ the emissivity of hydrogen ionising photons
emissivity is large enough that the post-overlap state of the
reionisation process should have been reached and the neutral fraction
of hydrogen should be small. Sources like A1835\#1916 may thus well be
responsible for the large electron scattering optical depth measured
by WMAP.  If the magnification is ${\cal M} \sim 5$ and/or the space
density is overestimated the star formation rate density at $z=10$
could be consistent with a smooth continuation of the observed
evolution at lower redshift.  These sources, however, would not be
sufficient to reionise at $z=10$ and the low neutral fraction inferred
by the detection of the \lya line would have to be explained in
another way, for instance by a fainter population of galaxies or by
partial ionisation by X-rays produced by accretion onto intermediate
mass black holes in mini-AGNs \citep{MadauR:03, RicottiO:03}.

The observed \Lya flux gives independent constraints on the ionisation
state of the surrounding IGM.  If the surrounding IGM were not
ionised, the strength of the line requires that its centre is
redshifted by 650 km s\m with respect to the velocity of the neutral
IGM. This could occur due to the absorption and resonant scattering of
the \Lya photons by a galactic wind, but such a large offset appears
unlikely considering the rather shallow potential well that may host
this galaxy. The constraints tighten further if the observed
correlation between transmission and velocity shift of \Lya emission
in LBGs also holds for A1835\#1916.  In this case the minimum size of
the ionised region consistent with a neutral surrounding IGM is
$R_{\rm S} \sim 5$ Mpc (comoving), independent of the velocity shift.

We agree with \cite{Loeb:04} that the source itself is unlikely to
ionise such a large region on its own.  However, the large emissivity
of ionising photons which is implied by the small effective survey
volume, if confirmed, would make the lack of a suppression of the \Lya
emission due to the Gunn-Peterson absorption by the surrounding IGM
less surprising.  Clearly, a small neutral fraction at $z=10$ in the
diffuse IGM would be further good news for surveys of high-redshift
objects which would strongly benefit from \Lya emitters with large
equivalent widths.  A space density as large as inferred here would
also mean that a significant fraction of the faintest objects detected
at $1.6 \mu {\rm m}$ may be at $z\sim 10$.

\section*{Acknowledgements.} 
M. Ricotti is supported by a PPARC theory grant. This work was
partially supported by the EC RTN network ``The Physics of the IGM''.
We thank Andy Bunker and Elizabeth Stanway for providing the data
compilation used to produce Fig.~1. We thank Roser Pello for pointing
out a mistake in our calculation of the comoving volume in an earlier
version of the manuscript.

\bibliographystyle{/home/ricotti/Latex/TeX/apj}
\bibliography{/home/ricotti/Latex/TeX/archive}

\begin{thebibliography}{}

\bibitem[\protect\citeauthoryear{{Barkana} \& {Loeb}}{{Barkana} \&
  {Loeb}}{2000}]{Barkana:00}
{Barkana}, R.,  \& {Loeb}, A. 2000, \apj, 539, 20

\bibitem[\protect\citeauthoryear{{Bouwens}, {Broadhurst}, \&
  {Illingworth}}{{Bouwens} et~al.}{2003}]{BouwensBI:03}
{Bouwens}, R., {Broadhurst}, T.,  \& {Illingworth}, G. 2003, \apj, 593, 640

\bibitem[\protect\citeauthoryear{{Bouwens} et~al.}{{Bouwens}
  et~al.}{2003}]{Bouwens:03}
{Bouwens}, R.~J., et~al. 2003, \apj, 595, 589

\bibitem[\protect\citeauthoryear{{Bouwens} et~al.}{{Bouwens}
  et~al.}{2004}]{Bouwens:04}
{Bouwens}, R.~J., et~al. 2004, ArXiv Astrophysics e-prints (astro-ph/0403167)

\bibitem[\protect\citeauthoryear{{Bromm}, {Kudritzki}, \& {Loeb}}{{Bromm}
  et~al.}{2001}]{Bromm:01}
{Bromm}, V., {Kudritzki}, R.~P.,  \& {Loeb}, A. 2001, \apj, 552, 464

\bibitem[\protect\citeauthoryear{{Bunker} et~al.}{{Bunker}
  et~al.}{2004}]{Bunker:04}
{Bunker}, A.~J., {Stanway}, E.~R., {Ellis}, R.~S.,  \& {McMahon}, R.~G. 2004,
  ArXiv Astrophysics e-prints (astro-ph/0403223)

\bibitem[\protect\citeauthoryear{{Cen}}{{Cen}}{2003}]{Cen:03a}
{Cen}, R. 2003, \apj, 591, 12

\bibitem[\protect\citeauthoryear{{Ciardi}, {Ferrara}, \& {White}}{{Ciardi}
  et~al.}{2003}]{CiardiFW:03}
{Ciardi}, B., {Ferrara}, A.,  \& {White}, S.~D.~M. 2003, \mnras, 344, L7

\bibitem[\protect\citeauthoryear{{Connolly} et~al.}{{Connolly}
  et~al.}{1997}]{Connolly:97}
{Connolly}, A.~J., {Szalay}, A.~S., {Dickinson}, M., {Subbarao}, M.~U.,  \&
  {Brunner}, R.~J. 1997, \apjl, 486, L11

\bibitem[\protect\citeauthoryear{{Efstathiou}}{{Efstathiou}}{1992}]{Efstathiou%
:92b}
{Efstathiou}, G. 1992, \mnras, 256, 43P

\bibitem[\protect\citeauthoryear{{Fan} et~al.}{{Fan} et~al.}{2003}]{Fan:03}
{Fan}, X., et~al. 2003, \aj, 125, 1649

\bibitem[\protect\citeauthoryear{{Fontana} et~al.}{{Fontana}
  et~al.}{2003}]{Fontana:03}
{Fontana}, A., {Poli}, F., {Menci}, N., {Nonino}, M., {Giallongo}, E.,
  {Cristiani}, S.,  \& {D'Odorico}, S. 2003, \apj, 587, 544

\bibitem[\protect\citeauthoryear{{Giavalisco} et~al.}{{Giavalisco}
  et~al.}{2004}]{Giavalisco:04}
{Giavalisco}, M., et~al. 2004, \apjl, 600, L103

\bibitem[\protect\citeauthoryear{{Guhathakurta}, {Tyson}, \&
  {Majewski}}{{Guhathakurta} et~al.}{1990}]{Guhathakurta:90}
{Guhathakurta}, P., {Tyson}, J.~A.,  \& {Majewski}, S.~R. 1990, \apjl, 357, L9

\bibitem[\protect\citeauthoryear{{Haiman}}{{Haiman}}{2002}]{Haiman:02}
{Haiman}, Z. 2002, \apjl, 576, L1

\bibitem[\protect\citeauthoryear{{Haiman}, {Abel}, \& {Madau}}{{Haiman}
  et~al.}{2001}]{Haiman:01}
{Haiman}, Z., {Abel}, T.,  \& {Madau}, P. 2001, \apj, 551, 599

\bibitem[\protect\citeauthoryear{{Hu} et~al.}{{Hu} et~al.}{2004}]{Hu:04}
{Hu}, E.~M., {Cowie}, L.~L., {Capak}, P., {McMahon}, R.~G., {Hayashino}, T.,
  \& {Komiyama}, Y. 2004, \aj, 127, 563

\bibitem[\protect\citeauthoryear{{Iwata} et~al.}{{Iwata}
  et~al.}{2003}]{Iwata:03}
{Iwata}, I., {Ohta}, K., {Tamura}, N., {Ando}, M., {Wada}, S., {Watanabe}, C.,
  {Akiyama}, M.,  \& {Aoki}, K. 2003, \pasj, 55, 415

\bibitem[\protect\citeauthoryear{{Kennicutt}}{{Kennicutt}}{1998}]{Kennicutt:98}
{Kennicutt}, R.~C. 1998, \araa, 36, 189

\bibitem[\protect\citeauthoryear{{Kodaira} et~al.}{{Kodaira}
  et~al.}{2003}]{Kodaira:03}
{Kodaira}, K., et~al. 2003, \pasj, 55, L17

\bibitem[\protect\citeauthoryear{{Kogut} et~al.}{{Kogut}
  et~al.}{2003}]{Kogut:03}
{Kogut}, A., et~al. 2003, \apjs, 148, 161

\bibitem[\protect\citeauthoryear{{Lehnert} \& {Bremer}}{{Lehnert} \&
  {Bremer}}{2003}]{Lehnert:03}
{Lehnert}, M.~D.,  \& {Bremer}, M. 2003, \apj, 593, 630

\bibitem[\protect\citeauthoryear{{Lilly} et~al.}{{Lilly}
  et~al.}{1996}]{Lilly:96}
{Lilly}, S.~J., {Le Fevre}, O., {Hammer}, F.,  \& {Crampton}, D. 1996, \apjl,
  460, L1

\bibitem[\protect\citeauthoryear{{Loeb}, {Barkana}, \& {Hernquist}}{{Loeb}
  et~al.}{2004}]{Loeb:04}
{Loeb}, A., {Barkana}, R.,  \& {Hernquist}, L. 2004, ArXiv Astrophysics
  e-prints (astro-ph/0403193)

\bibitem[\protect\citeauthoryear{{Madau}, {Haardt}, \& {Rees}}{{Madau}
  et~al.}{1999}]{MadauHR:99}
{Madau}, P., {Haardt}, F.,  \& {Rees}, M.~J. 1999, \apj, 514, 648

\bibitem[\protect\citeauthoryear{{Madau} et~al.}{{Madau}
  et~al.}{2004}]{MadauR:03}
{Madau}, P., {Rees}, M.~J., {Volonteri}, M., {Haardt}, F.,  \& {Oh}, S.~P.
  2004, \apj, 604, 484

\bibitem[\protect\citeauthoryear{{Miralda-Escud{\' e}}}{{Miralda-Escud{\'
  e}}}{2003}]{Miralda:03b}
{Miralda-Escud{\' e}}, J. 2003, \apj, 597, 66

\bibitem[\protect\citeauthoryear{{Miralda-Escude}}{{Miralda-Escude}}{1998}]{Mi%
ralda:98}
{Miralda-Escude}, J. 1998, \apj, 501, 15

\bibitem[\protect\citeauthoryear{{Miralda-Escud\'e}, {Haehnelt}, \&
  {Rees}}{{Miralda-Escud\'e} et~al.}{2000}]{Miralda:00}
{Miralda-Escud\'e}, J., {Haehnelt}, M.,  \& {Rees}, M.~J. 2000, \apj, 530, 1

\bibitem[\protect\citeauthoryear{{Pell\'o} et~al.}{{Pell\'o}
  et~al.}{2004}]{Pello:04}
{Pell\'o}, R., {Schaerer}, D., {Richard}, J., {Borgne}, J.~.~L.,  \& {Kneib},
  J.~. 2004, \AA, 416, L35

\bibitem[\protect\citeauthoryear{{Rhoads} et~al.}{{Rhoads}
  et~al.}{2003}]{Rhoads:03}
{Rhoads}, J.~E., et~al. 2003, \aj, 125, 1006

\bibitem[\protect\citeauthoryear{{Ricotti}}{{Ricotti}}{2002}]{Ricotti:02}
{Ricotti}, M. 2002, \mnras, 336, L33

\bibitem[\protect\citeauthoryear{{Ricotti}, {Gnedin}, \& {Shull}}{{Ricotti}
  et~al.}{2002}]{RicottiGSb:02}
{Ricotti}, M., {Gnedin}, N.~Y.,  \& {Shull}, J.~M. 2002, \apj, 575, 49

\bibitem[\protect\citeauthoryear{{Ricotti} \& {Ostriker}}{{Ricotti} \&
  {Ostriker}}{2003}]{RicottiO:03}
{Ricotti}, M.,  \& {Ostriker}, J.~P. 2003, \mnras, in press (astro-ph/0311003)
  (paper~IIa)

\bibitem[\protect\citeauthoryear{{Ricotti} \& {Ostriker}}{{Ricotti} \&
  {Ostriker}}{2004}]{RicottiOI:03}
{Ricotti}, M.,  \& {Ostriker}, J.~P. 2004, \mnras, 350, 539 (paper~I)

\bibitem[\protect\citeauthoryear{{Santos}}{{Santos}}{2003}]{Santos:03}
{Santos}, M.~R. 2003, ArXiv Astrophysics e-prints (astro-ph/0308196)

\bibitem[\protect\citeauthoryear{{Santos} et~al.}{{Santos}
  et~al.}{2003}]{SantosE:03}
{Santos}, M.~R., {Ellis}, R.~S., {Kneib}, J., {Richard}, J.,  \& {Kuijken}, K.
  2003, ArXiv Astrophysics e-prints (astro-ph/0310478)

\bibitem[\protect\citeauthoryear{{Schaerer}}{{Schaerer}}{2003}]{Schaerer:03}
{Schaerer}, D. 2003, \aap, 397, 527

\bibitem[\protect\citeauthoryear{{Shapley} et~al.}{{Shapley}
  et~al.}{2003}]{Shapley:03}
{Shapley}, A.~E., {Steidel}, C.~C., {Pettini}, M.,  \& {Adelberger}, K.~L.
  2003, \apj, 588, 65

\bibitem[\protect\citeauthoryear{{Sokasian} et~al.}{{Sokasian}
  et~al.}{2004}]{Sokasian:03}
{Sokasian}, A., {Yoshida}, N., {Abel}, T., {Hernquist}, L.,  \& {Springel}, V.
  2004, \mnras, 350, 47

\bibitem[\protect\citeauthoryear{{Songaila}}{{Songaila}}{2004}]{Songaila:04}
{Songaila}, A. 2004, ArXiv Astrophysics e-prints (astro-ph/0402347)

\bibitem[\protect\citeauthoryear{{Stanway} et~al.}{{Stanway}
  et~al.}{2004}]{Stanway:04}
{Stanway}, E.~R., et~al. 2004, \apjl, 604, L13

\bibitem[\protect\citeauthoryear{{Steidel} et~al.}{{Steidel}
  et~al.}{1999}]{Steidel:99}
{Steidel}, C.~C., {Adelberger}, K.~L., {Giavalisco}, M., {Dickinson}, M.,  \&
  {Pettini}, M. 1999, \apj, 519, 1

\bibitem[\protect\citeauthoryear{{Steidel} \& {Hamilton}}{{Steidel} \&
  {Hamilton}}{1992}]{Steidel:92}
{Steidel}, C.~C.,  \& {Hamilton}, D. 1992, \aj, 104, 941

\bibitem[\protect\citeauthoryear{{Thompson} et~al.}{{Thompson}
  et~al.}{1999}]{Thompson:99}
{Thompson}, R.~I., {Storrie-Lombardi}, L.~J., {Weymann}, R.~J., {Rieke}, M.~J.,
  {Schneider}, G., {Stobie}, E.,  \& {Lytle}, D. 1999, \aj, 117, 17

\bibitem[\protect\citeauthoryear{{Tumlinson} \& {Shull}}{{Tumlinson} \&
  {Shull}}{2000}]{Tumlinson:00}
{Tumlinson}, J.,  \& {Shull}, J.~M. 2000, \apjl, 528, L65

\bibitem[\protect\citeauthoryear{{Wyithe} \& {Loeb}}{{Wyithe} \&
  {Loeb}}{2003}]{WyitheL:03}
{Wyithe}, J.~S.~B.,  \& {Loeb}, A. 2003, \apjl, 588, L69

\bibitem[\protect\citeauthoryear{{Yan} et~al.}{{Yan} et~al.}{1998}]{Yan:98}
{Yan}, L., {McCarthy}, P.~J., {Storrie-Lombardi}, L.~J.,  \& {Weymann}, R.~J.
  1998, \apjl, 503, L19

\end{thebibliography}

\end{document}